\documentclass[english]{spie} 
\usepackage[T1]{fontenc}
\usepackage[latin1]{inputenc} 
\usepackage{graphicx}
\def\d#1{{\rm d}\kern -0.2mm#1}
\def\eq{Eq}

\def\fig{Fig}
\def\figs{Figs}

\title{Advances in Modeling of Scanning Charged-Particle-Microscopy Images}

\author{Petr Cizmar, Andr\'as E. Vlad\'ar, and Michael T. Postek \skiplinehalf
National Institute of Standards and Technology (NIST)
\footnote{~~Contribution of the National Institute of Standards and Technology;
not subject to copyright. Certain commercial equipment is identified
in this report to adequately describe the experimental procedure.
Such identification does not imply recommendation or endorsement by
the National Institute of Standards and Technology, nor does it imply
that the equipment identified is necessarily the best available for
the purpose.}
, 100 Bureau Drive,
Gaithersburg, MD 20899, USA}

\begin{document} 
\maketitle

\begin{abstract} 
Modeling artificial scanning electron microscope (SEM) and scanning ion microscope images
has recently become important. This is because of the need to provide repeatable
images with a priori determined parameters. Modeled artificial images are highly
useful in the evaluation of new imaging and metrological techniques, like
image-sharpness calculation, or drift-corrected image composition (DCIC).
Originally, the NIST-developed artificial image generator was designed only to
produce the
SEM images of gold-on-carbon resolution sample for image-sharpness evaluation.
Since then, the new improved version of the software was written in C++
programming language and is in the Public Domain. The current
version of the software can generate arbitrary samples, any drift function, and
many other features. This work describes scanning in charged-particle
microscopes, which is applied both in the artificial image generator and the
DCIC technique. As an example, the performance of the DCIC technique is
demonstrated.
\end{abstract}

\section{Introduction}
Computational scanning electron microscopy\cite{postek-simmod} through rapid
artificial image modeling is gaining importance. It is a useful tool for evaluation of imaging and
metrology methods, since real SEMs or other charged-particle microscopes
cannot always provide repeatable images. For example, it is virtually impossible to obtain two real
SEM images that only differ in random noise. This is usually caused by many
perturbing factors like drift, sample charging, or electro-magnetic fields. The artificial
image generator is capable of modeling all important effects in a deterministic
way. One can a priori choose the drift function, the type
\cite{box-gaussgen,knuth-art2} and magnitude and type of noise, the
charged-particle-beam profile, etc. That being the case, computer generated artificial images may be
input to the imaging and metrological techniques and the results compared to the
chosen parameters, hence indicating the performance of given techniques. None of
these is possible with the real images, where these effects are present there,
but all are random and often even unknown. 

An advanced version of the artificial SEM image generator
\cite{cizmar-simim-scanning,cizmar-optimization-scanning} has been released as a public-domain software. It
is implemented as a library written in C++. This also allows for linking with
programs written in many other programming languages.  The software works in
Linux, Mac~OSX, Windows, and very probably in other UN*X systems as well, however,
the latter has not yet been tested. For faster and easier designing of
calculations, Lua\cite{lua-book} scripting was implemented. Lua is a scripting language
originally designed for data-entry applications. These days it is mostly
employed in computer games. It is one of the simplest and fastest scripting languages
available. A simple graphical user interface (GUI) has been written mainly for 
demonstration. One can very easily generate images of two types;
gold-on-carbon resolution sample and periodic semiconductor cross structures. The GUI
depends on wxWidgets\cite{wx-book} library which is multiplatform as well. 

One of the techniques that have been tested with modeled images is the
drift-corrected image composition (DCIC)\cite{cizmar-fastsum-arxiv},
which outputs significantly more accurate images than the traditional imaging
techniques. This is necessary for sub-nanometer-scale metrology, since the
conventional ``slow-scan'' and ``fast-scan'' techniques provide images that are
often distorted or blurry. The DCIC works with frames that are taken as
quickly as the capabilities of the instrument permit. Physical drift causes
displacement between each couple of frames. This displacement is searched for
with cross-correlation. Since the quickly acquired frames are usually
extensively noisy, a noise reduction is a part of the DCIC technique.

\section{Drift Distortion}
In the scanning microscopes, the image is formed by scanning across the sample
in a raster pattern.
Intensity value is acquired at each location on the sample. In digital scanning
microscopes, that corresponds
with a pixel in the image. The intensity value $\xi(\vec{r})$ depends on the
landing position of the electron beam $\vec{r}$.  Most SEMs use the raster
pattern. Let the raster pattern be defined by the time-dependent vector
function:
\begin{eqnarray}
\vec r_r(t) &=& M\left(x(t)\vec e_x+y(t)\vec e_y\right),\\
t_p &=& t_D + t_d,\nonumber\\
y(t) &=& \left\lfloor\frac{t}{Xt_p+t_j}\right\rfloor,\\
x(t) &=&\left\lfloor\frac{t}{t_p}\right\rfloor - Xy(t),\\
0 \le &t& \le Y(Xt_p+t_j),\nonumber
\label{fast_composition_eq_raster}
\end{eqnarray}
where $t$ is time, $M$ is a single-pixel step length. $x$ and $y$ are column and
row indexes in the SEM image.  $\vec{e_x}$ and $\vec{e_y}$ are the unit vectors
in x- and y-direction, $t_D$ is the pixel-dwell time, $t_d$ is the dead
time between acquisition of two pixels, $t_j$ is the time needed to move the beam to the
beginning of the new line. $\lfloor q \rfloor$ is a symbol for the ${\rm
floor}(q)$ function as used in programming languages. $X$ and $Y$ are the
pixel-width and pixel-height of the SEM image.
 
Let the SEM imaging be defined as a relation between the intensity map of the sample
$\xi(\vec{r})$ and the SEM image $I(x,y)$:
\begin{equation}
I(x(t),y(t)) = K\xi(\vec{r}(t)).
\label{fast_composition_eq_imaging}
\end{equation}
The relation between $I$ and $\xi$ may in practice be very general. For simplicity, let $K$
be a constant in this manuscript, since this does not affect generality of the
DCIC technique.
In the ideal case: $\vec{r}(t) = \vec r_r(t)$; however, drift and space
distortions are always present in scanning microscopes and
they can significantly affect the position $\vec{r}$:
\begin{equation}
\vec{r}(t) = \vec{r}_r(t) + \vec{D}_d(t) + \vec{D}_s(\vec{r}_r).
\label{fast_composition_eq_distortions}
\end{equation}
The space distortion $\vec D_s$ is constant in time and may be 
compensated for, when its function is known. This distortion may be caused by
non-linearities in deflection amplifiers and is significant mostly at
low magnifications. On the other hand, the drift distortion $\vec D_d$ is
changing in time, its function is usually unknown, and it may extensively 
affect the high-magnification images. The drift distortion may arise from several
sources; e.g. translational motion of the sample, tilt or deformation of the
electron-optical column, outer forces and vibrations, or temperature expansion.
High-magnification images are very sensitive to drift distortion, since
microscopic displacements, tilts, or temperature changes can easily cause
nanometer distortions and displacements, which can significantly impair the SEM image
and its usability for nanometer-scale measurements.

\begin{figure}[htb]
\begin{center}
\includegraphics[width=\textwidth]{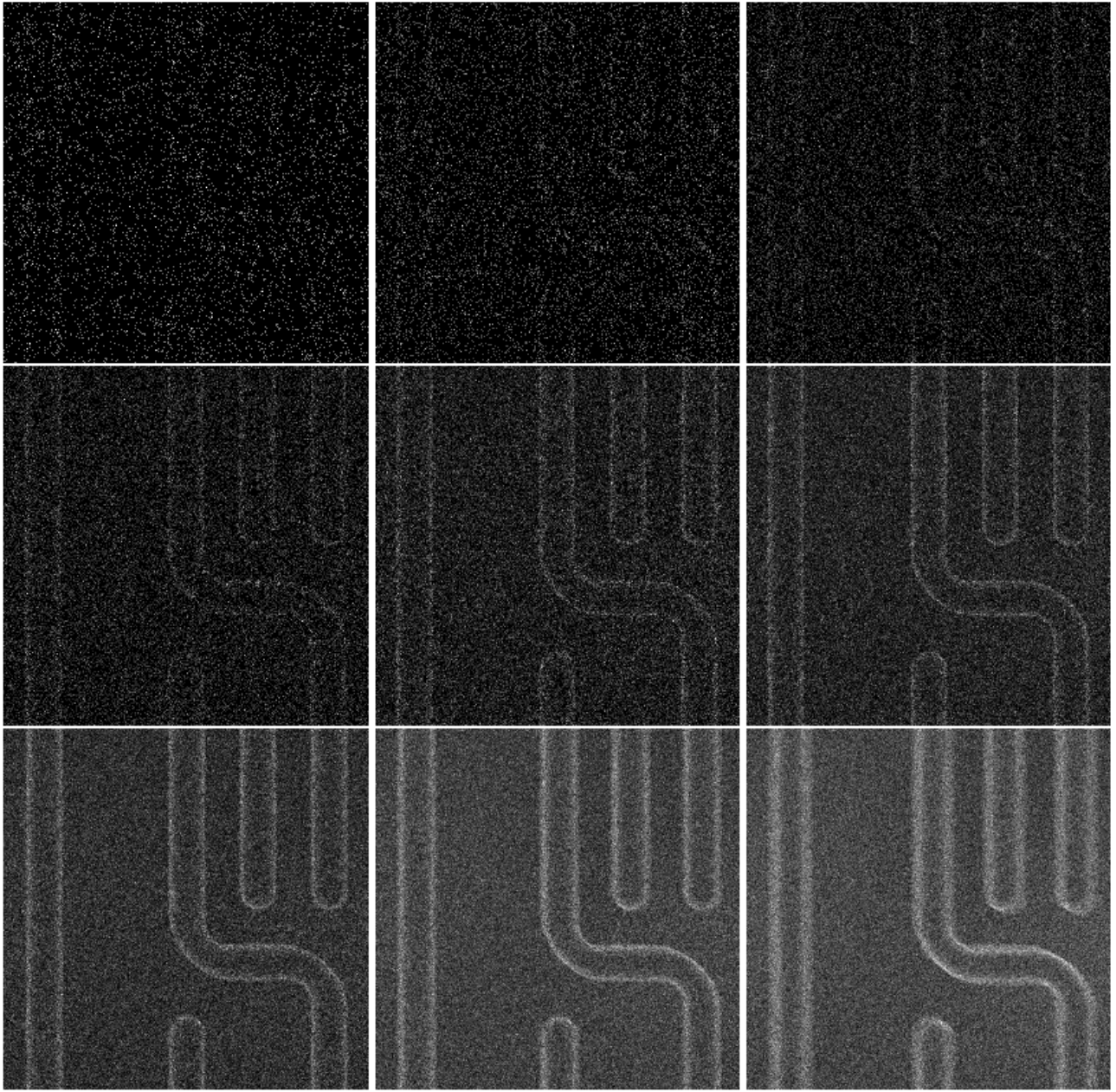}
\end{center}
\caption{Series of artificial images of a semiconductor structure composed using
the traditional ``fast-scan'' technique. Compositions of 2, 4, 8, 16, 32, 64,
128, 256, and 512 frames (from the top left). Images are normalized.}
\label{fast-scan-series}
\end{figure}

\begin{figure}[htb]
\begin{center}
\includegraphics[width=\textwidth]{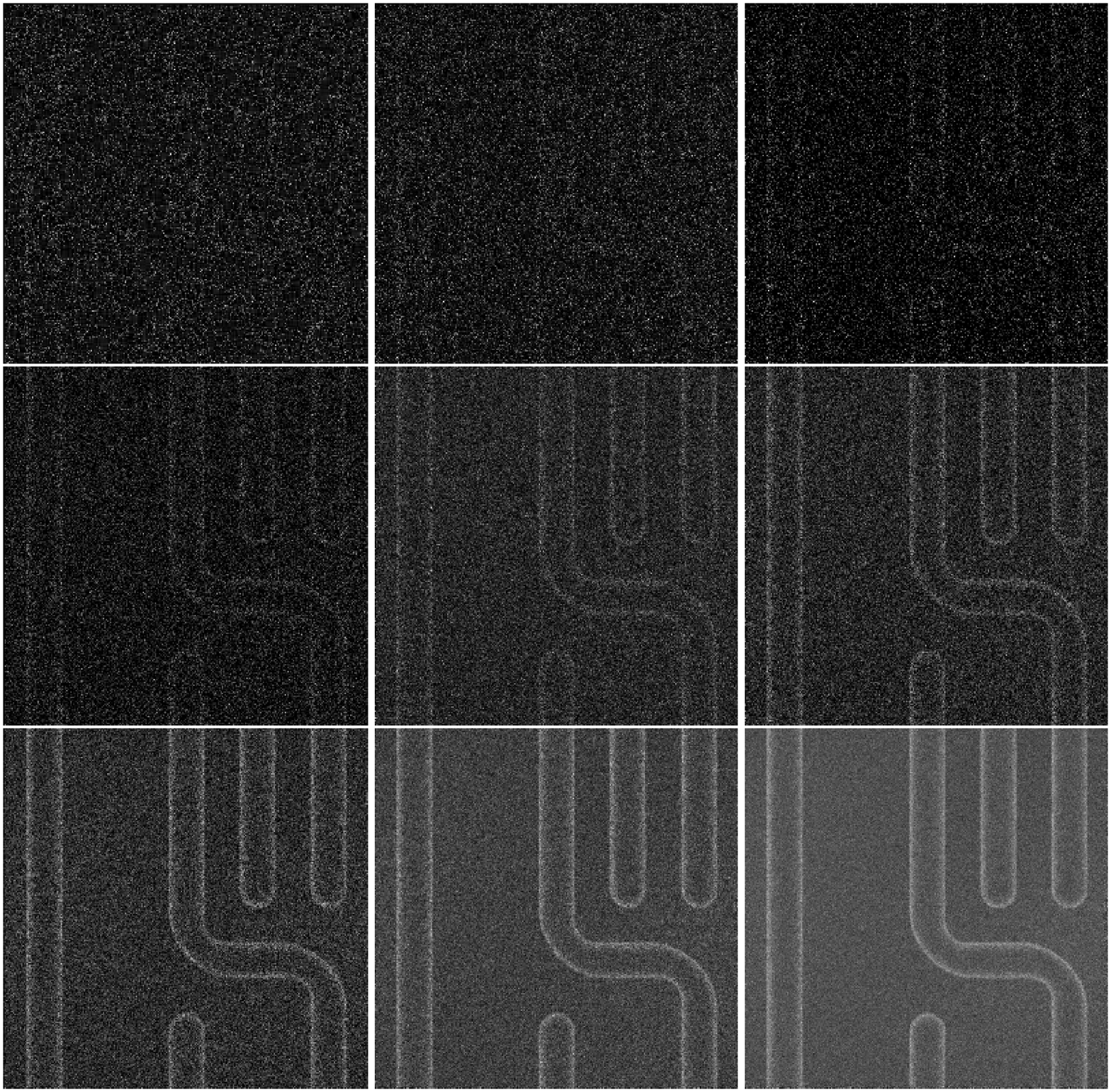}
\end{center}
\caption{Series of artificial images of a semiconductor structure composed using
the DCIC method. Compositions of 2, 4, 8, 16, 32, 64, 128, 256, and 512 frames
(from the top left).}
\label{dcic-series}
\end{figure}

The drift-distortion function is generally unknown, however, since it
characterizes motion of physical bodies, it must be continuous and thus
square-integrable. Therefore, drift-distortion function may be Fourier-series
expanded:
\begin{eqnarray}
D_{cd}(t) &=& \sum\limits_{n=-\infty}^\infty c_n{\rm e}^{-{\rm i}nt},\\
\vec{D}_d &=& \Re(D_{cd}) \vec{e}_x + \Im(D_{cd}) \vec{e}_y,\\ 
U &\propto& \sum_{n=-\infty}^\infty c_n^2 n^2,
\label{fast_composition_eq_fourexp}
\end{eqnarray}
where $c_n$ are the (complex) Fourier coefficients, $U$ is the overall energy of the
drifting system.
Since $U$ is limited, for high $n$ the coefficients $c_n$ must be nearing zero.
In practice, for frequencies higher than 200~Hz, $c_n$ correspond to noise only
and are negligible. Therefore, the $D_{cd}(t)$ can be written:
\begin{equation}
D_{cd}(t) \approx \sum\limits_{n=-N}^N c_n{\rm e}^{-{\rm i}nt},\\
\label{fast_composition_eq_fourexplim}
\end{equation}
where $N$ represents the highest significant angular frequency.

\section{``Fast-scan'' Imaging}
The imaged intensity signal in the SEM always contains noise. The intensity
function is a superposition of a real signal and noise:
\begin{equation}
\xi(\vec{r},t) = \xi_s(\vec{r}) + \xi_n(t),
\label{fast_composition_eq_inten}
\end{equation}
where $\xi_s$ is the position-dependent real signal and $\xi_n$ is the
time-dependent noise. $\xi_n$ is a superposition of all noise contributions
present in the SEM:  Poisson noise originating
from the electron source and the secondary emission, the noise originating from
the amplifier and electronics, quantization-error noise, etc. Due to the central limit
theorem, it is legitimate to suppose that the mean value of this noise is zero:
\begin{equation}
<\xi_n(t)> = 0.
\label{fast_composition_eq_clt}
\end{equation}

In order to obtain a SEM image with a desired level of noise, the overall pixel
dwell-time $t_{D}$ must be sufficiently high.  Unfortunately, the electron yield
is usually low and the overall pixel-dwell time must often be set to times
ranging from tens to several hundreds of $\mu$s.

In the SEM, there are two
common methods to achieve this, i.e ``slow-scan'' and ``fast scan'', while the
latter is useful for metrological application.

``Fast-scan'' is one of the common imaging methods in SEMs. The image is
composed from multiple ($N$) frames, for which averaging is the mostly applied
technique. The
frames are acquired with the lowest possible pixel-dwell time $t_D$.  The image pixel
value is an average of corresponding frame-pixel values:
\begin{eqnarray}
I_k(x(t_0),y(t_0)) &=& K \xi_s(\vec r(t_0+kt_f))+\nonumber\\
 &+& K\xi_n(t_0+kt_f), 
\label{fast_composition_eq_fastscani}\\
I(x,y) &=& \frac{1}{N}\sum_{k=0}^{N} I_k(x,y).\\
t_f &=& Y(Xt_p+t_j)+t_{jj},
\end{eqnarray}
$t_f$ is a time period between beginnings of acquisition of two following
frames, $t_{jj}$ is the dead time between the end of acquisition of one frame and
beginning of the next one. Considering \eq~(\ref{fast_composition_eq_clt}), the
higher $N$, the lower noise level is present in the image. The required noise-level
thus determines the number of composed images $N$. For high $N$:
\begin{equation}
\sum_{k=0}^{N-1} \xi_n(t_0+kt_f) \approx 0.\\
\end{equation}
Because the scanning raster pattern is constant for all frames,
\begin{equation}
\vec r_r(t_0+kt_f) = \vec r(t_0).\\
\end{equation}
\eq~(\ref{fast_composition_eq_fastscani}) may be expanded:
\begin{eqnarray}
I(x(t_0),y(t_0)) &=& \frac{K}{N} \sum_{k=0}^{N-1} \xi_s[\vec r_r(t_0) + 
\nonumber\\
&+& \vec D_s(\vec r_r(t_0)) + \vec D_d(t_0+kt_f)].
\label{fast_composition_eq_fastscani_exp}
\end{eqnarray}

With current SEMs, the frame-acquisition time $t_f$ can be much lower than
the period of even the highest drift-distortion frequencies. The
drift-distortion within the single-frame acquisition time is then minimal.
However, it becomes significant during acquisition of the whole image,
especially, when the dead times $t_{jj}$ are prohibitively high, which is the case
even with some current instruments.

\section{Drift-Corrected Image Composition (DCIC)}

The ``fast-scan'' method may be significantly improved using drift-distortion
correction, when the images are acquired quickly enough.  Since the space-distortion $\vec D_s$ is much less
pronounced and much smaller that the drift-distortion $\vec D_d$ at very high
magnifications, it will be neglected from now on. The
\eq~(\ref{fast_composition_eq_fastscani_exp}) then becomes:
\begin{eqnarray}
I(x,y) &=& \frac{K}{N} \sum_{k=0}^{N-1} \xi_s[\vec r_r(r) + \vec
D_{dk}],\\
\vec D_{dk} &=& \vec D_d(t_0+kt_f).
\end{eqnarray}
The image is in this case the mean value of $N$ displaced images. 

Under certain conditions, it is possible to find the displacement
vectors of the images, which are equal to the drift-distortion values $\vec D_{dk}$.
The drift-distortion then may be compensated for, which allows for acquisition of a
corrected, more accurate image.
One possible approach is a cross-correlation-based displacement
detection, which is used in the DCIC technique. The maximum of the
cross-correlation function is searched for. Its position is equal to the
searched displacement vector $\vec D_{dk}$.

In the DCIC technique, the cross-correlation with noise reduction is applied. This
is necessary, because the quickest-acquired images are usually very noisy and
the peak in the cross-correlation function becomes overridden by numerous other
peaks, corresponding to random correlation
of noise.  This often makes
finding the displacement vector impossible. This issue can be tackled by low-pass
frequency filtering performed in the frequency domain. The cut-off
frequency is determined by the filter-radius $R$.  

\begin{figure}[tb]
\begin{center}
\includegraphics{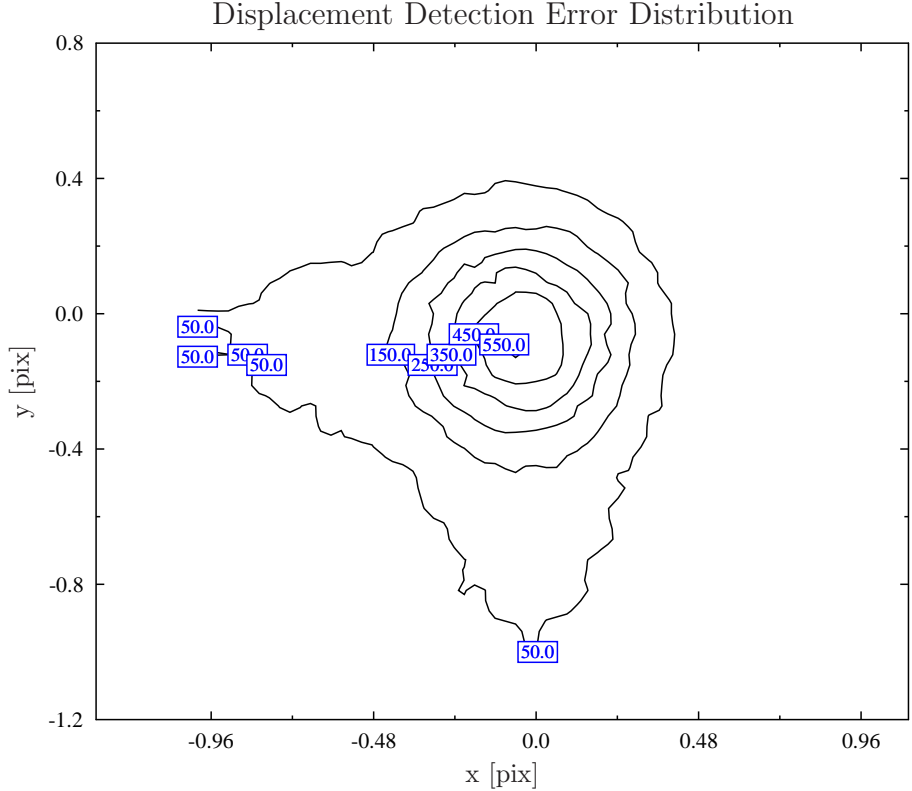}
\end{center}
\caption{Error distribution of the displacement detection. Artificial SEM image
of a periodic semiconductor sample was used.}
\label{error-profile}
\end{figure}

Plain maximum search in a discrete function limits the accuracy to a minimum of
one pixel. 
However, in the DCIC, the detection of the displacement vector $\vec D_{dk}$ is performed with
sub-pixel resolution. The peak in the two-dimensional cross-correlation function
is interpolated with a polynomial third-order two-dimensional polynomial
function and the algorithm then searches for its maximum. 

The technique is very powerful, since it can correct for the drift-related
distortions and blur in extremely noisy images. (See \figs~\ref{fast-scan-series} and \ref{dcic-series}) 

\section{Accuracy of the DCIC technique}
The accuracy of the detected displacement vector $\vec D_{dk}$ characterizes the
accuracy of the DCIC imaging technique. Errors in the displacement vector can
cause blur. Such blur can under certain circumstances be larger than with
application of the
original ``fast-scan'' technique ($\vec D_{dk} = \vec 0$). In metrological
applications, where dimensions are measured from the images, the drift-related
displacement is the main source of errors. 

The artificial SEM images have been successfuly used to evaluate accuracy of the
DCIC technique. The artificial-image generator is, unlike any other source of
SEM images, capable of modeling all necessary characteristics for this
application, e.g. arbitrary drift functions, dead times, arbitrary types of
samples, etc. 

A performance characteristics must be chosen to investigate the limits
of an imaging technique. Application of the standard deviation of the 
displacement vector would be a good candidate, if the distribution of errors was
Gaussian. 
In order to find this out, a large set of artificial images (500\,000) randomly differing in
displacement and noise has been applied to find the error distribution of the
displacement detection. The DCIC technique has processed all generated frames
and has output corresponding displacement values.  The two-dimensional histogram of
these values forms the resulting distribution, which is shown in the
\fig~\ref{error-profile}. These data have clearly indicated that the error
distribution is not (always) Gaussian. Using standard (Gaussian) error
processing has therefore been unsuitable and thus we have chosen the mean error
$\delta_D$ as the performance characteristics.
\begin{equation}
\bar\delta_D = \frac{1}{N-1} \sum_{k=1}^{N-1} \delta_{Dk},
\label{deltad}
\end{equation}
where $\delta_{Dk}$ is the error of the displacement vector $\vec D_{dk}$ and
$N$ is the number of frames. Since the correct displacement vector  $\vec
D_{ci}$ is known (it is determined by the artificial-image generator), 
\begin{equation}
\delta_{Dk}= |\vec D_{dk} - \vec D_{ck}|.
\label{deltadi}
\end{equation}

The performance of the DCIC technique is obviously limited, because noise, blur,
contrast, and other parameters affect it significantly. For instance, if the
frames were extremely blurred and the cross-correlation maximum would be overly
wide and the mean error of the displacement vector would be excessively high. It
is therefore useful to find the dependences of $\delta_{D}$ on noise and blur
and provide a set of limiting parameters.

\begin{figure}[t]
\begin{center}
\includegraphics{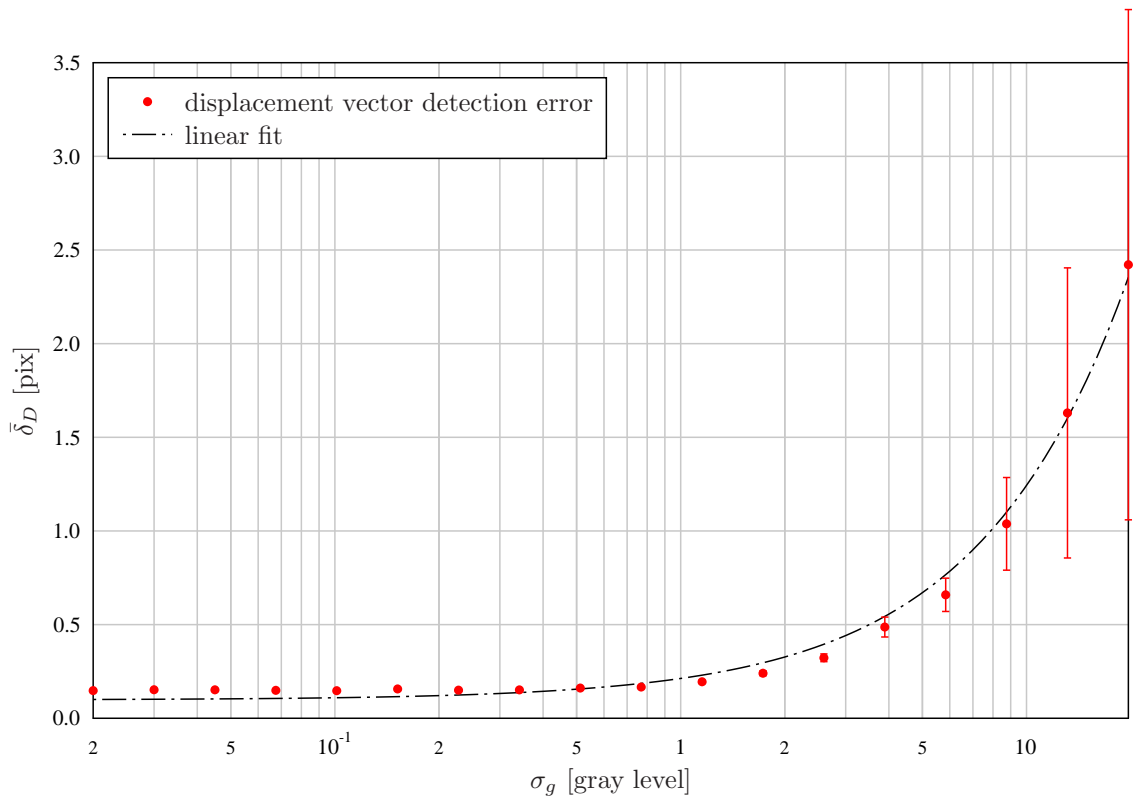}
\end{center}
\caption{Evaluation of the DCIC technique. Dependence of the mean displacement
detection error on the magnitude ($\sigma_g$) of Gaussian noise. Each represents
5000 artificial images of the gold-on-carbon resolution sample sized 512x512
pixels. The error-bars denote the standard deviation of the displacement vector detection error.}
\label{noise-test}
\end{figure}

\begin{figure}[t]
\begin{center}
\begin{tabular}{cccc}
\hline
$\sigma_g = 10^{-2}$ &
$\sigma_g = 10^{-1}$ &
$\sigma_g = 1$ &
$\sigma_g = 10$ \\
$SNR = 60$ &
$SNR = 6$ &
$SNR = 0.6$ &
$SNR = 0.06$ \\
$SNR_{dB} = 17.8{\rm ~dB}$ &
$SNR_{dB} = 7.78{\rm ~dB}$ &
$SNR_{dB} = -2.22{\rm ~dB}$ &
$SNR_{dB} = -12.2{\rm ~dB}$ \\
\hline\noalign{\smallskip}
\includegraphics[width=0.20\textwidth]{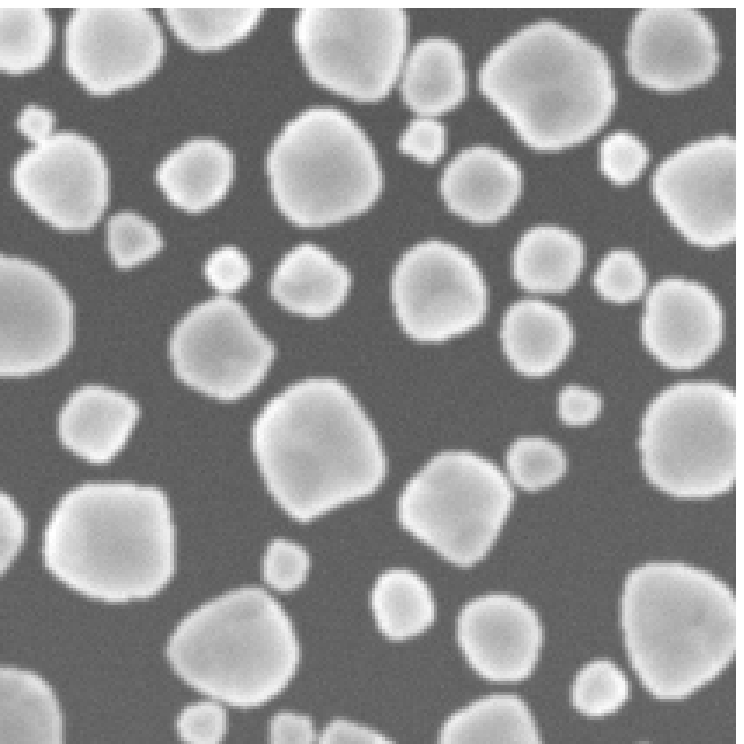} & 
\includegraphics[width=0.20\textwidth]{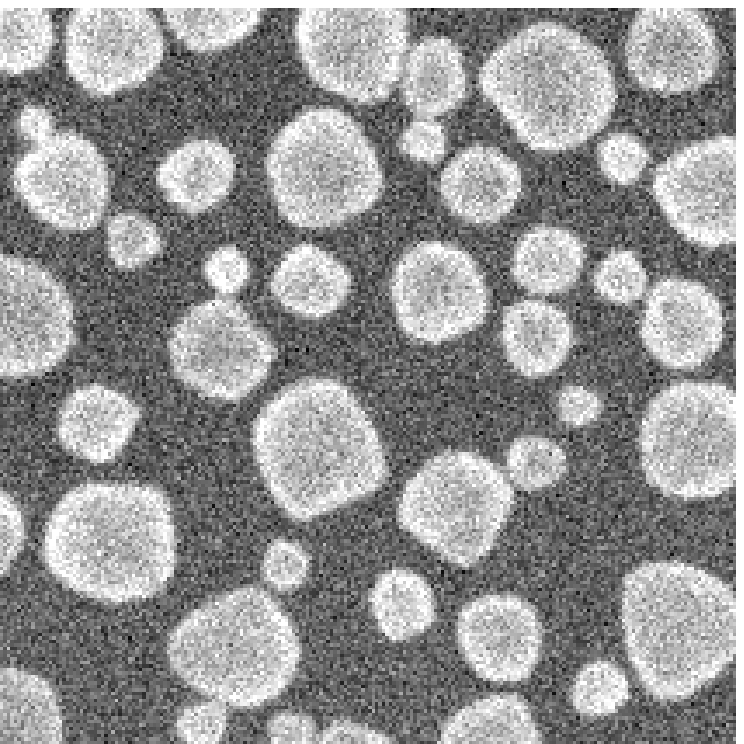} & 
\includegraphics[width=0.20\textwidth]{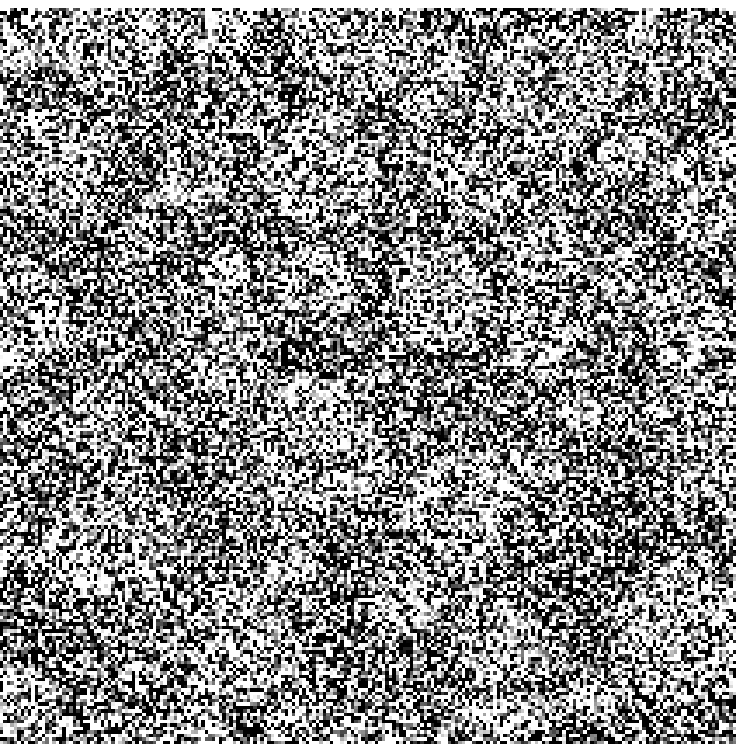} & 
\includegraphics[width=0.20\textwidth]{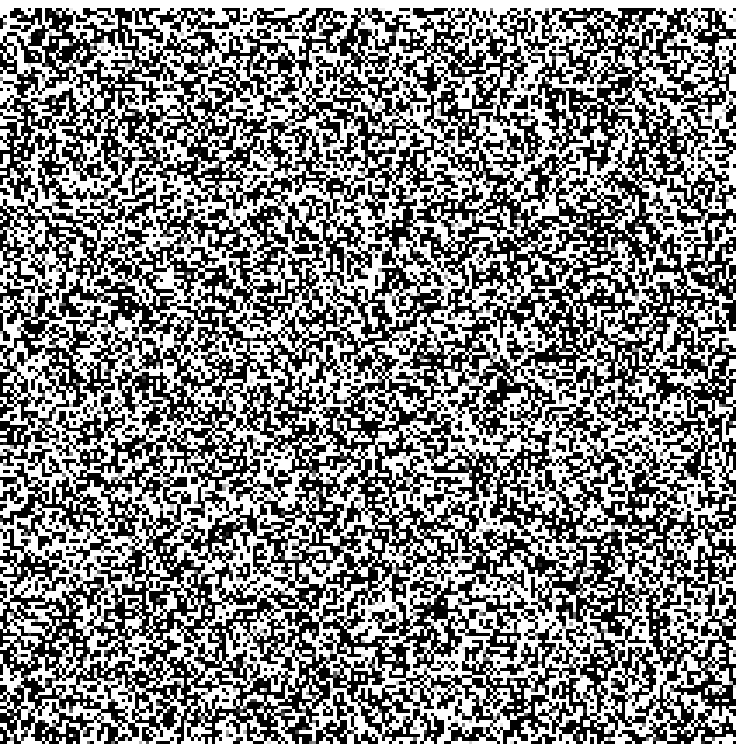} \\
\hline
\end{tabular}
\end{center}
\caption{Gaussian-noise scale. Artificial images of the gold-on-carbon
resolution sample with Gaussian noise of different magnitudes.}
\label{noise-scale}
\end{figure}

\begin{figure}[tb]
\begin{center}
\includegraphics{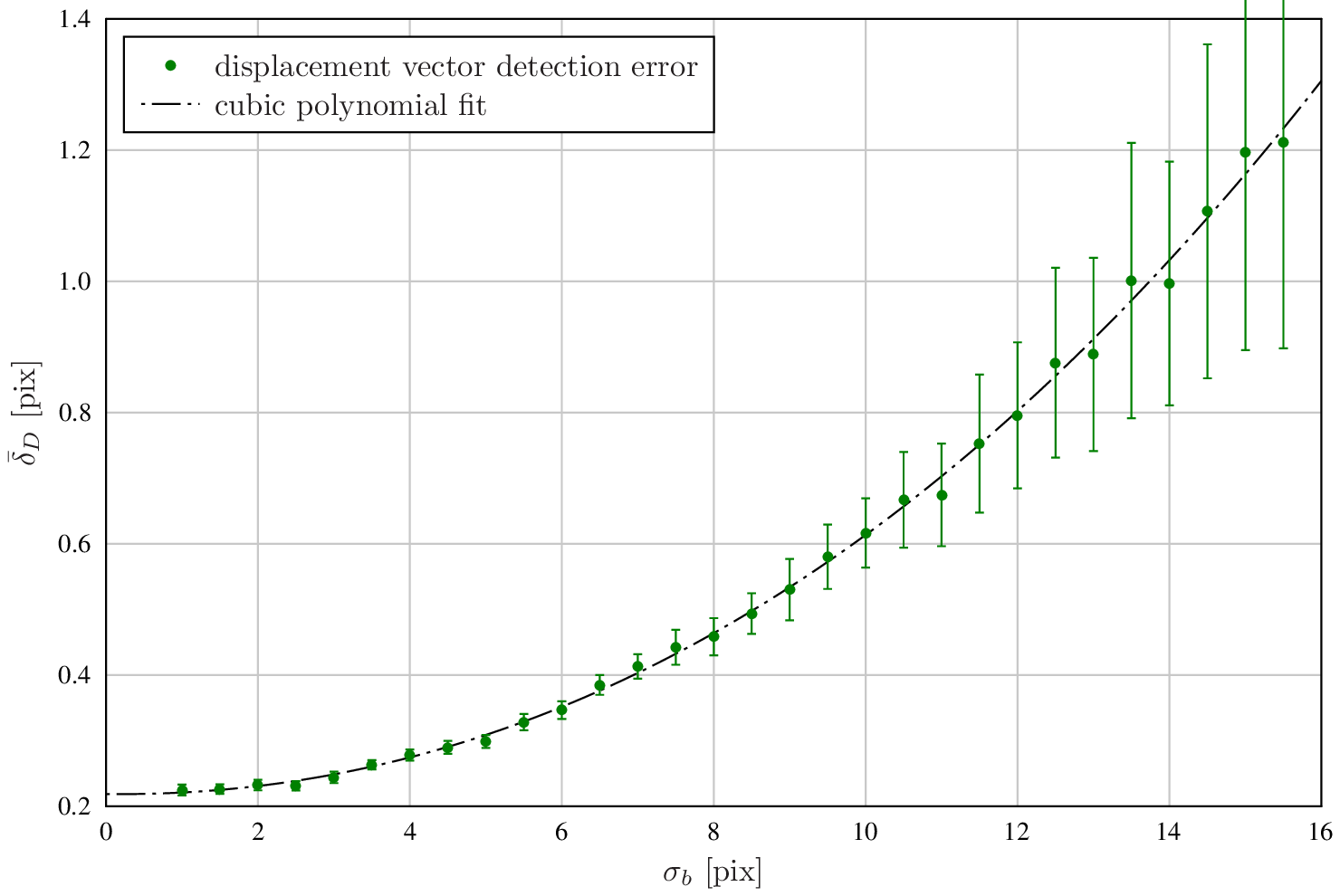}
\end{center}
\caption{Evaluation of the DCIC technique. Dependence of the mean displacement
detection error on Gaussian blur ($\sigma_b$). This blur simulates the effect of
the charged-particle-beam profile.  Each represents 5000 artificial images of
the gold-on-carbon resolution sample sized 512x512 pixels. The error-bars denote
the standard deviation of the displacement vector detection error.}
\label{blur-test}
\end{figure}

The dependence of the mean error of the detected displacement $\bar \delta_D$ on
noise and blur have been both investigated with application of artificial
images. Gaussian noise and Gaussian blur have been chosen for simplicity,
although the type of noise and the blur profile may be arbitrary. For every step
in noise and blur, 5000 artificial images of the gold-on-carbon resolution
sample have been generated and processed by the DCIC algorithm. The results of
these tests are shown in \figs~\ref{noise-test} and \ref{blur-test}. For
reference images showing different magnitudes of Gaussian noise see
\fig~\ref{noise-scale}. These tests demonstrate the capability of the DCIC
technique to find the displacements with sub-pixel accuracy. In the noise test,
this is maintained up to the $\sigma_g = 8$, which roughly corresponds to signal
to noise ratio (SNR) around 0.1 and the dependence is almost linear. The
dependence on blur indicates that the sub-pixel accuracy is
sustained up to $\sigma_b = 14$. 

\section{Conclusion}
Modeled artificial SEM images were first employed in assessment of the
image-sharpness calculation techniques\cite{postek-simmod} and have been adopted
as a part of the developed
international standard for image sharpness. Since then, a new highly improved version of the
software was written. This version supports arbitrary non-overlapping two-dimensional
samples, rigorous generation of Poisson and Gaussian noise, arbitrary drift
functions, dead times and other features. Scripting in Lua scripting language
was implemented to make the calculations easier to design. This new tool
was then used in evaluation of the new imaging technique of DCIC. By finding
dependence of the error in detection of the displacement on noise and blur, the
sub-pixel accuracy was demonstrated even for high magnitudes of noise or blur.
This makes the DCIC and modeling of microscope images useful and important tools
for nanoscale metrology and nanotechnology.

\bibliography{literature}{} 
\bibliographystyle{spiebib} 

\end{document}